\documentclass[10.5pt,a4paper,oneside]{extarticle}
\usepackage[T1]{fontenc} 
\usepackage[left=25mm,right=20.9mm,top=3.75cm,bottom=3.25cm,headsep=0.25cm,footskip=0.25cm]{geometry}

\usepackage[english]{babel}
    \usepackage[affil-it]{authblk}
    \usepackage{etoolbox}
    \usepackage{lmodern}
    \usepackage{setspace}
    \usepackage{amssymb,amsthm,amsmath}
    \usepackage{tikz}
    \usepackage{listings}
    \usepackage{color}
    \usepackage{caption}
    \usepackage{placeins}
    \usepackage{booktabs}
    \usepackage{enumerate}
    \usepackage{graphicx}
    \usepackage{titlesec}
    \usepackage{mathptmx}
    \usepackage{anyfontsize}
    \usepackage{t1enc}
    \usepackage[utf8]{inputenc}
    \usepackage{fancyhdr}
    \usepackage{float,xspace}
    \usepackage{scalefnt}
	
\usepackage{soul} 
\usepackage{tabulary}
\usepackage[colorlinks=true,linkcolor=blue]{hyperref} 

\usepackage{textcase}

\usepackage[colorinlistoftodos]{todonotes}

\def\defeq{\stackrel{\triangle}{=}}
\usepackage[caption=false,font=normalsize,labelfont=sf,textfont=sf]{subfig}

\titlespacing*{\section}
{0pt}{12pt}{0pt}
\titlespacing*{\subsection}
{0pt}{12pt}{0pt}
\titlespacing*{\subsubsection}
{0pt}{12pt}{0pt}
\titleformat{\section}
  {\huge\sffamily\Large\bfseries\color{black}}
  {\thesection}{1em}{}
    \titleformat{\subsection}[block]{\bfseries}{\thesubsection}{.5em}{}
    \titleformat{\subsubsection}[block]{\bfseries}{\thesubsubsection}{.5em}{}

\titleformat{\section}{\fontsize{12}{19}\bfseries}{\thesection}{1em}{}

\usepackage{mathptmx} 

\newtheorem{myDef}{Property}

\usepackage{newtxtext}
\usepackage{newtxmath}

\makeatletter

\patchcmd{\@maketitle}{\LARGE \@title}{\fontsize{14}{19.2}\selectfont\@title}{}{} 
\makeatother
\title
{
	\vspace{-5cm}
	\begin{minipage}{\textwidth}	
	\hspace{-20pt}\includegraphics[width=9.5cm]{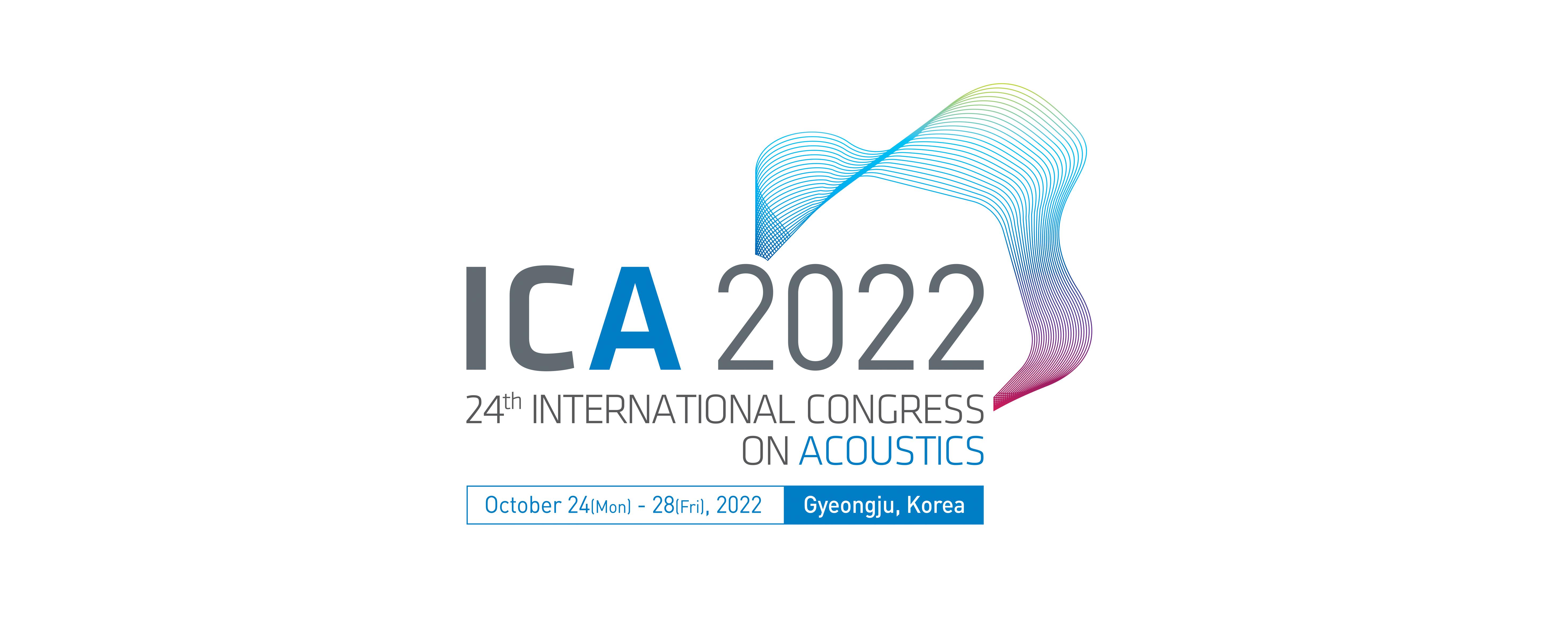}
	\hspace{-65pt}\includegraphics[width=10.2cm]{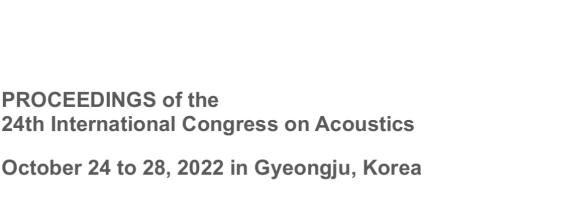}
	\end{minipage}
	\includegraphics[width=16.5cm]{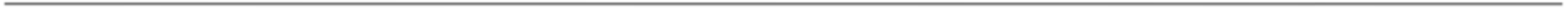}\\[0.5cm] \textbf{Independent low-rank matrix analysis based on the Sinkhorn divergence source model for blind source separation}
	\author[ ]{Jianyu Wang$^{(1)}$, Shanzheng Guan$^{(1)}$, Jingdong Chen$^{(1)}$, and Jacob Benesty$^{(2)}$}
  	\affil[(1)]{Center of Intelligent Acoustics and Immersive Communication, Northwestern Polytechnical University, Xi’an, 710072, China alexwang96@mail.nwpu.edu.cn, gshanzheng@mail.nwpu.edu.cn, jingdongchen@ieee.org}
    \affil[(2)]{INRS-EMT, University of Quebec, 800 de la Gauchetiere Ouest, Suite 6900, Montreal, QC H5A 1K6, Canada jacob.benesty@inrs.ca}
}
\date{}

\begin{document}

\clearpage
\setcounter{page}{1}
\maketitle
\thispagestyle{empty}
\fancypagestyle{empty}
{	
	\fancyhf{} \fancyfoot[R]
	{
		\vspace{-2cm}\includegraphics[width=17cm]{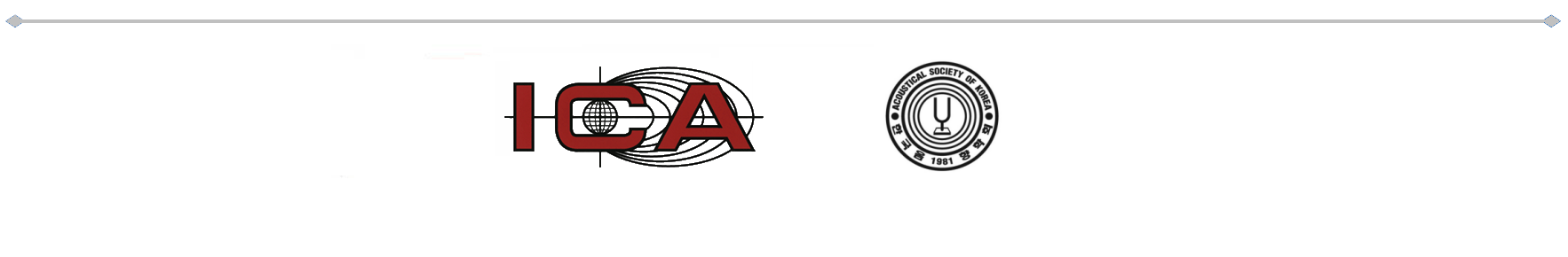}
	}
}

\subsection*{\fontsize{10.5}{19.2}\uppercase{\textbf{Abstract}}} \vskip 6pt
{\fontsize{10.5}{60}\selectfont The so-called independent low-rank matrix analysis (ILRMA) has demonstrated a great potential for dealing with the problem of determined blind source separation (BSS) for audio and speech signals. This method assumes that the spectra from different frequency bands are independent and the spectral coefficients in any frequency band are Gaussian distributed. The Itakura-Saito divergence is then employed to estimate the source model related parameters. In reality, however, the spectral coefficients from different frequency bands may be dependent, which is not considered in the existing ILRMA algorithm. This paper presents an improved version of ILRMA, which considers the dependency between the spectral coefficients from different frequency bands. The Sinkhorn divergence is then exploited to optimize the source model parameters. As a result of using the cross-band information, the BSS performance is improved. But the number of parameters to be estimated also increases significantly, and so is the computational complexity. To reduce the algorithm complexity, we apply the Kronecker product to decompose the modeling matrix into the product of a number of matrices of much smaller dimensionality. An efficient algorithm is then developed to implement the Sinkhorn divergence based BSS algorithm and the complexity is reduced by an order of magnitude.}

\noindent{\\ \fontsize{11}{60}\selectfont Keywords: Independent low-rank matrix analysis (ILRMA), Blind source separation (BSS), Sinkhorn distance, Kronecker product.} 

\fontdimen2\font=4pt

\section{\uppercase{Introduction}} \vskip 6pt

Multichannel blind source separation (BSS) refers to the problem of estimating source signals from their mixtures observed by an array of sensors without using any prior information about the mixing system \cite{ref1}. For audio and speech applications \cite{ref2}, the problem can be divided into two cases: underdetermined and determined. The former refers to the case where the number of sensors in the array is less than the number of sources. In this case, the problem cannot be solved without additional information or constraints \cite{ref3, ref4}. The latter refers to the scenario where the number of sensors is greater than or equal to the number of sources. In this case, separation can be achieved by identifying the demixing system from only the observation signals. This work focus on the latter case, i.e., the determined BSS for audio and speech signals.

In audio and speech applications, the signal observed at every sensor is a mixture of all the source signals convolved with the corresponding acoustic channel impulse responses. As the acoustic channel impulse responses are usually very long (it is not uncommon to have a few thousands of points),
this convolutive mixing process make it challenging and difficult to achieve source separation directly in the time domain from the perspectives of accuracy, robustness, and complexity. A widely adopted approach to circumventing this issue is to transform the time-domain signals into the time-frequency domain using the short-time fourier transform (STFT), thereby converting the convolutive mixing problem into one of instantaneous mixing. Consequently, majority of efforts in audio and speech BSS have been focused in the STFT domain. Many methods and algorithms have been developed in this domain over the last few decades and the representative ones include the so-called independent component analysis (ICA) \cite{ref6} and independent vector analysis (IVA) \cite{ref7, ref8}. In comparison, IVA based methods are more appropriate than ICA for dealing with audio BSS in the STFT domain as it dramatically mitigates the permutation problem. While they have demonstrated reasonably good performance, the classical IVA algorithms do not take advantage of the structural information in the source spectra, which are useful to improve BSS performance. To exploit such information, Daichi {\em et al.} proposed an independent-low-rank-matrix-analysis (ILRMA) method \cite{ref5}, which utilizes nonnegative matrix factorization (NMF) to decompose the given spectrogram as the product between basis and temporal activation matrices. By assuming that the spectral components from different frequency bands are independent and the spectral coefficients in any frequency band are Gaussian distributed, this method employs the Kullback-Leibler (KL) or Itakura-Saito (IS) divergence as the cost function to estimate the parameters of the NMF-based source model.

However, the spectral components of the same source from different frequency bands may be correlated as demonstrated in the literature of noise reduction \cite{benesty2011,huang2014}, which is not considered in the ILRMA algorithm. This paper presents an improved version of ILRMA, which takes advantage of the cross-band dependency of spectra to improve BSS performance. We adopt the Sinkhorn divergence \cite{ref9}, \cite{ref11}, \cite{ref12} as the cost function to optimize the parameters of the NMF-based source model, resulting in a Sinkhorn divergence based ILRMA (SDILRMA) algorithm. 
Since the cross-band information is used, SDILRMA is able to improve the BSS performance. But the number of parameters to be estimated also increases significantly, and so is the computational complexity. To reduce the number of parameters and the algorithm complexity, we subsequently apply the Kronecker product tool \cite{ref13, ref14} to decompose the modeling matrix into the product of a number of matrices of much smaller dimensionality, leading to a simplified SDILRMA, which is computationally more efficient than its original counterpart and is able produce better performance than ILRMA.

\section{\uppercase{SIGNAL MODEL AND PROBLEM FORMULATION}} \vskip 6pt

Suppose that there are $N$ sources in the sound field and we use a microphone array consisting of $M$ sensors to pick up the signals. The observation signal at the $m$th microphone and time index $j$ is then
\begin{align}\label{eq1}
  x_m(j) &= \sum_{n=1}^N a_{nm}(j) * s_{n}(j) , 
\end{align}
where $s_n(j)$ denotes the $n$th source signal and $a_{nm}(j)$ is the acoustic impulse response from the $n$th source to the $m$th sensor.

Transforming both sides of \eqref{eq1} into the short-time Fourier transform (STFT) domain and rearranging the results into a vector form gives
\begin{align}\label{eq3}
\begin{split}
  \mathbf{x}_{f,t} &= \sum_{n=1}^N \mathbf{a}_{n,f} S_{n,f,t} \\
                   &= \sum_{n=1}^N \mathbf{x}_{n,f,t} ,
\end{split}
\end{align}
where $S_{n,f,t}$ is the STFT of $s_{n}(j)$, $\mathbf{x}_{f,t} \defeq \left[ X_{1,f,t}, ~\cdots, ~X_{M,f,t} \right]^T \in \mathbb{C}^M$ with $X_{m,f,t}$ being the STFT of $x_m(j)$,  $\mathbf{a}_{n,f} \defeq \left[ A_{n,1,f},~\cdots,~A_{n,M,f} \right]^T$ with $A_{n,m,f}$ denoting the acoustic transfer function, the superscript $^T$ denotes the transpose operator, $f$ and $t$ denote, respectively, the frequency and frame indices, and $\mathbf{x}_{n,f,t} \defeq \mathbf{a}_{n,f} S_{n,f,t}$, whose elements are often called the source images.

The signal model in \eqref{eq3} can be rearranged into a more compact form as
\begin{align}\label{model2}
  \mathbf{x}_{f,t} = \mathbf{A}_f \mathbf{s}_{f,t},
\end{align}
where $\mathbf{A}_f \defeq \left[ \mathbf{a}_{1,f},~\cdots,~\mathbf{a}_{n,f} \right] \in \mathbb{C}^{M\times N}$ is called the mixing matrix, and $\mathbf{s}_{f,t} \defeq \left[ S_{1,f,t}, ~\cdots, ~S_{N,f,t} \right]^T$ is a vector consisting of the $N$ source signals.  Now, the problem of BSS becomes one of identifying a demixing matrix such that
\begin{align}\label{separate}
  \mathbf{y}_{f,t} = \mathbf{D}_f \mathbf{x}_{f,t},
\end{align}
where  $\mathbf{D}_f = \left[ \mathbf{d}_{1,f},~\cdots,~\mathbf{d}_{N,f} \right] \in \mathbb{C}^{N\times M}$ denotes the demixing matrix, and $\mathbf{y}_{f,t}$ is an estimate of $\mathbf{s}_{f,t}$ (up to a scale and permutation). Note that if the mixing matrix $\mathbf{A}_f = \left[ \mathbf{a}_{1,f},\dots,\mathbf{a}_{n,f} \right] \in \mathbb{C}^{M\times N}$ is not singular as assumed in such methods as ILRMA, the demixing matrix should be the inverse of the mixing matrix $\mathbf{A}_f$.

To achieve this identification, some source model has to be assumed. The so-called spherically invariant random processing (SIRP) model has been widely used in BSS for speech signals  \cite{ref17}. With this model, the multivariate probability density function can be derived from the corresponding univariate probability density function and the correlation matrices \cite{ref16,ref18}. As a particular case of SIRP, the local Gaussian model has gained much attention, in which the source spectrum in every time-frequency (TF) bin is modeled as a time-varying complex Gaussian distribution \cite{ref15} and the spectral components from different frequency bins and time frames are assumed to be mutually independent, and as a result, $s_{n,f,t}$ follows a zero-mean complex Gaussian distribution with a time-varying variance $\lambda_{n,f,t}$, i.e.,
\begin{equation}\label{eq4}
\begin{split}
  s_{n,f,t} \sim \mathcal{N}_{\mathbb{C}}\left( 0,\lambda_{n,f,t} \right).
\end{split}
\end{equation}
The critical parameter of this source model is the time-varying variance $\lambda_{n,f,t}$, which needs to be estimated. One way to achieve such estimation is through NMF, in which the variance matrix of every source is modeled as a low-rank approximation of the product of a basis matrix and an activation matrix. Given $\lambda_{n,f,t}$, the variance matrix is defined as
\begin{align}
\label{nmbdam}
\boldsymbol{\lambda}_{n} \defeq
\begin{bmatrix}
\lambda_{n,1,1} & \dots & \lambda_{n,1,T}\\
\vdots &\ddots & \vdots \\
\lambda_{n,F,1}& \dots & \lambda_{n,F,T}
\end{bmatrix}, \quad
\end{align}
which consists of the time-varying variance for all the time frames (the total number of frames is denoted as $T$) and frequencies bins (the number of frequency bins is denoted as $F$). The the low-rank approximation is then expressed as 
\begin{align}
\label{eq5}
\boldsymbol{\lambda}_{n} \approx \mathbf{W}_n \mathbf{H}_n, 
\end{align}
where 
\begin{align}
\mathbf{W}_n &=
\begin{bmatrix}
w_{n,1,1} & \dots & w_{n,1,K}\\
\vdots &\ddots & \vdots \\
w_{n,F,1}& \dots & w_{n,F,K}
\end{bmatrix}, \quad \\
   \mathbf{H}_n &=
\begin{bmatrix}
h_{n,1,1} & \dots & h_{n,1,T}\\
\vdots &\ddots & \vdots \\
h_{n,K,1}& \dots & h_{n,K,T}
\end{bmatrix},
\end{align}
are, respectively, the basis and activation matrices, and $K$ denotes the number of basis vectors. With this approximation, the estimation of the time-varying variances, i.e.,  $\lambda_{n,f,t}$, for all the time frames and frequency bins is converted to a problem of estimating the basis and activation matrices, which will be discussed in the next section.

From \eqref{eq3} and \eqref{eq4}, one can check that $\mathbf{x}_{n,f,t}$ follows a multivariate complex Gaussian distribution, i,e,.
\begin{equation}\label{eq6}
  \mathbf{x}_{n,f,t} \sim \mathcal{N}_{\mathbb{C}} \left( \mathbf{0},  ~~\lambda_{n,f,t}\mathbf{R}_{n,f} \right),
\end{equation}
where $\mathbf{0}$ is column vector with all its elements being 0,  $\mathbf{R}_{n,f} \defeq E \left[ \mathbf{x}_{n,f,t} \mathbf{x}_{n,f,t}^H\right] $ is the spatial covariance matrix for the $n$th source. If one approximates this matrix as $\mathbf{R}_{n,f} = \mathbf{a}_{n,f}\mathbf{a}_{n,f}^H$, the model degenerates to a rank-1 spatial model. Given $\mathbf{R}_{n,f}$, one can be check that the observation signal vector  $\mathbf{x}_{f,t}$ follows the following distribution:
\begin{equation}\label{eq7}
\begin{split}
\mathbf{x}_{f,t} \sim \mathcal{N}_{\mathbb{C}} \left(\mathbf{0}, ~~\sum_{n=1}^N \lambda_{n,f,t}\mathbf{R}_{n,f} \right).
\end{split}
\end{equation} 

\section{SINKHORN DIVERGENCE BASED MODEL PARAMETER ESTIMATION} \vskip 6pt

Generally, the NMF based source model adopts the IS divergence as the cost function for optimization. For the ILRMA algorithm, the cost function, which is denoted as $\mathcal{L}_{\mbox{ILRMA}}$, is the sum of the logarithmic conditional probability $p\left( \mathbf{X}_{f,t}|\lambda_{n,f,t},\mathbf{D}_f \right)$, i.e., 
\begin{align}\label{eq9}
  & \mathcal{L}_{\mbox{ILRMA}} = \sum_{f=1}^{F} \sum_{t=1}^{T} \log\left[ p\left(\left. \mathbf{X}_{f,t} \right| \lambda_{n,f,t},\mathbf{D}_f \right) \right] \nonumber \\
  &= \sum_{f=1}^F \sum_{t=1}^{T} \log \mathcal{N}_{\mathbb{C}} \left( \mathbf{X}_{f,t} \left| \mathbf{0}, \sum_{n=1}^N \lambda_{n,f,t} \mathbf{a}_{n,f} \mathbf{a}_{n,f}^H \right.\right) \nonumber \\
  & = -\sum_{f=1}^{F} \sum_{t=1}^{T} \mbox{Tr} \left[ \mathbf{y}_{f,t}^H \mathbf{D}_{f}^{-H} \left( \mathbf{D}_f^H \boldsymbol{\Lambda}_{f,t}^{-1} \mathbf{D}_f \right) \mathbf{D}_f^{-1} \mathbf{y}_{f,t} \right] + T\sum_{f=1}^{F} \log \left| \mathbf{D}_f \mathbf{D}_f^H \right| - \sum_{f=1}^{F} \sum_{t=1}^{T} \log\left| \boldsymbol{\Lambda}_{f,t} \right| + \mbox{Cst} \nonumber \\
  & = - \sum_{f=1}^{F} \sum_{t=1}^{T} \left[ \sum_{n=1}^{N} \frac{\left| y_{n,f,t} \right|^2}{\sum_{k=1}^{K}w_{n,f,k}h_{n,k,t}} + \sum_{n=1}^{N} \log \left(\sum_{k=1}^{K} w_{n,f,k} h_{n,k,t} \right) \right] + 2T\sum_{f=1}^{F} \log\left| \mathbf{D}_f \right| + \mbox{Cst},
\end{align}
where $\boldsymbol{\Lambda}_{f,t} = \mbox{Diag}\left( \lambda_{1,f,t},~\dots,~\lambda_{N,f,t} \right)$ is a diagonal matrix. 
Note that the first term on the right-hand side of the last line in \eqref{eq9} denotes the source model, which can also be viewed as the IS divergence between the low-rank approximated spectra and the estimated source spectra for every source, and the second term denotes the spatial model.

{
It is seen from \eqref{eq9} that the spectra from different frequency bins are treated independently. In practice, the spectral components of the same source from different frequency bins may be correlated \cite{benesty2011,huang2014}. In what follows, we introduce the Sinkhorn divergence based source model to replace the first term on the right-hand side of the last line in \eqref{eq9} so the cross-band information is used to estimate the model parameters. Specifically, the Sinkhorn divergence is expressed as
\begin{align}\label{eq10}
  D_{\mathrm{S}} \left( \mathbf{Y}_n \cdot \mathbf{Y}_n^* ~|~ \boldsymbol{\lambda}_n \right) = &\sum_{t=1}^T ~\min_{\mathbf{P}_t} ~\langle \mathbf{P}_t, \mathbf{C} \rangle - \frac{1}{\mu} H\left( \mathbf{P}_t \right)  ~~~\mbox{s. \ t.} \quad \mathbf{P}_t\mathbf{1} = \mathbf{y}_{n,t} \cdot \mathbf{y}_{n,t}^*, \quad \mathbf{P}_t \mathbf{1}^T = \boldsymbol{\lambda}_{n,t},
\end{align}
where $\langle \rangle$ denotes the inner product between two matrices, $\cdot$ denotes the Hadamard Product (element-wise Multiplication), $\mathbf{P}_t\in U\left(\mathbf{y}_{n,t} \cdot \mathbf{y}_{n,t}^*,  \boldsymbol{\lambda}_{n,t}  \right)$ denotes the transport matrix with $\left[ \mathbf{P}_t \right]_{ij}$ describing the frequency component migrates from the $i$th frequency bin of $\mathbf{y}_{n,t} \cdot \mathbf{y}_{n,t}^*$ to the $j$th subband of $\boldsymbol{\lambda}_{n,t}$, $\mathbf{1} \in \mathbb{R}^{F}$ is the all-one vector,  $U\left(\mathbf{y}_{n,t} \cdot \mathbf{y}_{n,t}^*,  \boldsymbol{\lambda}_{n,t} \right):=\left\{ \mathbf{P}_t \in \mathbb{R}_{+}^{F\times F} ~|~ \mathbf{P}_t \mathbf{1} = \mathbf{y}_{n,t} \cdot \mathbf{y}_{n,t}^*, ~\mathbf{P}_t^T \mathbf{1} =  \boldsymbol{\lambda}_{n,t} \right\}$ denotes a transport polytope, which contains all paths from the estimated source $\mathbf{y}_{n,t} \cdot \mathbf{y}_{n,t}^*$ to the target parameter $\boldsymbol{\lambda}_n$, $\mathbf{C}\in\mathbb{R}^{F\times F}$ represents the cost of transporting one unit of the source vector to the target vector, and
$H(\mathbf{P}_t) = -\sum_{i,j} \left[ \mathbf{P}_t \right]_{ij} \log \left[ \mathbf{P}_t \right]_{ij}$
denotes the entropic regularization term, which enables efficient approximation of the gradient of the Sinkhorn divergence. 

Using the Lagrange multiplier method, one can express \eqref{eq10} as 
\begin{equation}\label{eq11}
\begin{split}
  \!\!\!D_\mathrm{S}^{\mu,\gamma} \left(\left. \mathbf{Y}_n \cdot \mathbf{Y}_n^* ~\right|~ \boldsymbol{\lambda}_n \right) = \sum_{t=1}^{T} \left[ \min_{\mathbf{P}_t} ~\langle \mathbf{P}_t, ~\mathbf{C} \rangle - \frac{1}{\mu} H(\mathbf{P}_t) + \gamma D_{\mathrm{KL}} \left( \mathbf{P}_t \mathbf{1} \left|  \mathbf{y}_{n,t} \cdot \mathbf{y}_{n,t}^* \right. \right) + \gamma D_{\mathrm{KL}}\left( \left. \mathbf{P}_{t} \mathbf{1}^T \right| \boldsymbol{\lambda}_{n,t} \right) \right],
\end{split}
\end{equation}
where $\boldsymbol{\lambda}_{n,t} = \sum_{k=1}^K \mathbf{w}_{n,k} h_{n,k,t}$, and $D_{\mathrm{KL}} \left( x | y \right) = x\log\frac{x}{y} -x + y$. Note that only a single Lagrange multiplier is used in \eqref{eq11} to reduce the number of parameters.

The transport matrix $\mathbf{P}_t$ should satisfy $\mathbf{P}_t = \mbox{diag}(\mathbf{u})\ \mathbf{G} \ \mbox{diag}(\mathbf{v})$ when optimizing the cost function in \eqref{eq11}, where $\mathbf{u} = \left( \frac{\mathbf{y}_{n,t} \cdot \mathbf{y}_{n,t}^*}{\mathbf{P}_t \mathbf{1}} \right)^{\gamma\mu}$, $\mathbf{v}=\left( \frac{\boldsymbol{\lambda}_{n,t}}{\mathbf{P}_t \mathbf{1}^T} \right)^{\gamma\mu}$ (note that here the fraction between two vectors denotes the element wise division), and $\mathbf{G} = \exp(-\mu \mathbf{C} - 1)$. The optimal transport matrix $\mathbf{P}_t$ is estimated by a Sinkhorn-like iterative algorithm.

For the basis matrix $\mathbf{W}_n$ and the activation matrix $\mathbf{H}_n$, we construct an auxiliary function as
\begin{equation}\label{eq12}
\begin{split}
  A\left( \mathbf{W}_n, \mathbf{W}_n^\star \right) = \sum_{t=1}^{T} \sum_{k_1,\dots,k_F} \prod_{f} \alpha_{f,k_f} D_\mathrm{S}^{\mu,\gamma} \left( \mathbf{y}_{n,t} \cdot \mathbf{y}_{n,t}^* \left| \frac{\sum_{k=1}^{K} \mathbf{w}_{n,k} h_{n,k,t}}{\boldsymbol{\alpha}} \right. \right),
\end{split}
\end{equation}
\begin{equation}\label{eq13}
\begin{split}
  A\left( \mathbf{H}_n, \mathbf{H}_n^\star \right) = \sum_{t=1}^{T} \sum_{k_1,\dots,k_F} \prod_{f} \beta_{f,k_f} D_\mathrm{S}^{\mu,\gamma} \left( \mathbf{y}_{n,t} \cdot \mathbf{y}_{n,t}^* \left| \frac{\sum_{k=1}^{K} \mathbf{w}_{n,k} h_{n,k,t}}{\boldsymbol{\beta}} \right.\right)
\end{split}
\end{equation}
where $\mathbf{H}_n^\star$ denotes an auxiliary matrix constructed from $\mathbf{H}$, $\alpha_{f,k_f} = \frac{w_{n,f,k_f}^\star h_{n,k_f,t}}{\sum_{k_f} w_{n,f,k_f} h_{n,k_f,t}^\star}$, and $\beta_{f,k_f} = \frac{w_{n,f,k_f}h_{n,k_f,t}^\star}{\sum_{k_f} w_{n,f,k_f} h_{n,k_f,t}^\star}$. Through evaluating the partial derivatives $\frac{\partial A\left( \mathbf{W}_n, \mathbf{W}_n^\star \right)}{ \partial w_{n,f,k}}$ and $\frac{\partial A\left( \mathbf{H}_n, \mathbf{H}_n^\star \right)}{\partial h_{n,k,t}}$, we can obtain the algorithm to estimate the elements of the basis and activation matrices, i.e.,
\begin{equation}\label{eq14}
\begin{split}
  w_{n,f,k} \leftarrow w_{n,f,k} \sqrt{\frac{\sum_t \left[ \mathbf{P}_t \mathbf{1} \right]_f h_{n,k,t} \left( \sum_{k^\prime} w_{n,f,k^\prime} h_{n,k^\prime,t} \right)^{-2} }{\sum_{t} \left[ \mathbf{P}_t \mathbf{1} \right]_f \left( \sum_{k^\prime} w_{n,f,k^\prime} h_{n,k^\prime,t} \right)^{-1}}},
\end{split}
\end{equation}
\begin{equation}\label{eq15}
\begin{split}
  h_{n,k,t} \leftarrow h_{n,f,k} \sqrt{\frac{\sum_f \left[ \mathbf{P}_t \mathbf{1} \right]_f w_{n,f,k} \left( \sum_{k^\prime} w_{n,f,k^\prime} h_{n,k^\prime,t} \right)^{-2}}{\sum_f \left[ \mathbf{P}_t \mathbf{1} \right]_f \left( \sum_{k^\prime} w_{n,f,k^\prime} h_{n,k^\prime,t} \right)^{-1}}}.
\end{split}
\end{equation}

The model parameters are optimized in a similar manner as ILRMA \cite{ref5}. Note, however, computation of the transport matrix $\mathbf{P}_t$ in every frame for the  $n$th source requires large memory and is computationally expensive. In the next section, we apply the Kronecker product tool to decompose the transport matrix $\mathbf{P}_t$ into a product of a number of matrices of much smaller dimensionality.
}

\section{MODEL PARAMETER ESTIMATION BASED ON KRONECKER PRODUCT DECOMPOSITION} \vskip 6pt

\begin{myDef}{(sum of Kronecker product)\cite{ref13}:}
Let two matrices be $\mathbf{A} \in \mathbb{R}^{m\times m}$ and $\mathbf{B} \in \mathbb{R}^{n\times n}$, their Kronecker sum can be expressed as
\begin{equation}\label{eq16}
\begin{split}
  \mathbf{A} \oplus \mathbf{B} = \mathbf{A} \otimes \mathbf{I}_n + \mathbf{I}_m \otimes \mathbf{B},
\end{split}
\end{equation}
where $\mathbf{I}_m$ and $\mathbf{I}_n$ are identity matrices of size $m\times m$ and $n \times n$, respectively, and $\otimes$ denotes the Kronecker product.
\end{myDef}
Since the above Kronecker product decomposition is based on two all-one matrices, we name it the all-one Kronecker product.

{
Let us decompose the cost matrix $\mathbf{C}$ as
\begin{align}
\mathbf{C} = \oplus_{q=1}^Q \mathbf{C}_q = \mathbf{C}_1 \otimes \mathbf{C}_2 \otimes \cdots \otimes \mathbf{C}_Q,
\end{align}
where $\mathbf{C}_1 \in \mathbb{R}^{f_1\times f_1},\dots, \mathbf{C}_{Q} \in \mathbb{R}^{f_Q\times f_Q}$, $F=f_1\times \cdots \times f_Q$. The intermediate variable matrix $\mathbf{G}$ can then be written as
\begin{equation}\label{eq17}
\begin{split}
  \mathbf{G} = \exp\left( -\mu \oplus_{q=1}^Q \mathbf{C}_q - 1 \right) = e^{-1} \otimes_{q=1}^Q \exp\left( -\mu \mathbf{C}_q \right).
\end{split}
\end{equation}

The product $\mathbf{P}_t\mathbf{1}$ in \eqref{eq14} and \eqref{eq15} can be calculated in another way:
\begin{equation}\label{eq18}
\begin{split}
  \mathbf{P}_t \mathbf{1} = \mbox{diag} \left( \mathbf{u} \right) \mathbf{G} \mbox{diag} \left( \mathbf{v} \right)\mathbf{1} = \mbox{diag}\left( \mathbf{u} \right) \mathbf{G} \mathbf{v} = \mbox{diag}(\mathbf{u}) e^{-1} \otimes_{q=1}^Q \exp\left( -\mu \mathbf{C}_q \right) \mathbf{v}.
\end{split}
\end{equation}

Now, let us use the relationship between vector-operator and Kronecker product, i.e.,  $\mbox{vec}\left( \mathbf{ABC} \right) = \left( \mathbf{C}^T \otimes \mathbf{A} \right) \mbox{vec}\left( \mathbf{B} \right)$.  Then, we adopt a fold operator $\mbox{fold}\left( \cdot \right)$ and a product operator $\times_{q=1}^Q$ to fold a vector into a tensor, thereby transforming the vector $\mathbf{v} \in \mathbb{R}^F$ into an $Q$ order tensor $\mathcal{V} = \mbox{fold}(\mathbf{v}) \in \mathbb{R}^{f_1\times f_2 \times \dots \times f_Q}$. This gives
\begin{equation}\label{eq19}
\begin{split}
  P_t \mathbf{1} = \mbox{diag} \left( \mathbf{u} \right) \mbox{vec}\left( \mathcal{V} \times_{q=1}^Q \exp \left( -\mu \mathbf{C}_q \right) \right).
\end{split}
\end{equation}

Note that \eqref{eq19} does not require to compute directly the transport matrix $\mathbf{P}_t$, which helps reduce the computational complexity by a magnitude.
Now, the estimators in \eqref{eq14} and \eqref{eq15} can be updated as
\begin{equation}\label{eq20}
\begin{split}
  w_{n,f,k} \leftarrow w_{n,f,k} \sqrt{\frac{\sum_t \left[ \mbox{diag} \left( \mathbf{u} \right) \mbox{vec}\left( \mathcal{V} \times_{q=1}^Q \exp \left( -\mu \mathbf{C}_q \right) \right) \right]_f h_{n,k,t} \left( \sum_{k^\prime} w_{n,f,k^\prime} h_{n,k^\prime,t} \right)^{-2} }{\sum_{t} \left[ \mbox{diag} \left( \mathbf{u} \right) \mbox{vec}\left( \mathcal{V} \times_{q=1}^Q \exp \left( -\mu \mathbf{C}_q \right) \right) \right]_f \left( \sum_{k^\prime} w_{n,f,k^\prime} h_{n,k^\prime,t} \right)^{-1}}},
\end{split}
\end{equation}
\begin{equation}\label{eq21}
\begin{split}
  h_{n,k,t} \leftarrow h_{n,f,k} \sqrt{\frac{\sum_f \left[ \mbox{diag} \left( \mathbf{u} \right) \mbox{vec}\left( \mathcal{V} \times_{q=1}^Q \exp \left( -\mu \mathbf{C}_q \right) \right) \right]_f w_{n,f,k} \left( \sum_{k^\prime} w_{n,f,k^\prime} h_{n,k^\prime,t} \right)^{-2}}{\sum_f \left[ \mbox{diag} \left( \mathbf{u} \right) \mbox{vec}\left( \mathcal{V} \times_{q=1}^Q \exp \left( -\mu \mathbf{C}_q \right) \right) \right]_f \left( \sum_{k^\prime} w_{n,f,k^\prime} h_{n,k^\prime,t} \right)^{-1}}}.
\end{split}
\end{equation}
}

\begin{figure*}[t]
\centering
\vspace{-0.7cm} 
\subfloat[female+female]{\includegraphics[width=1.85in]{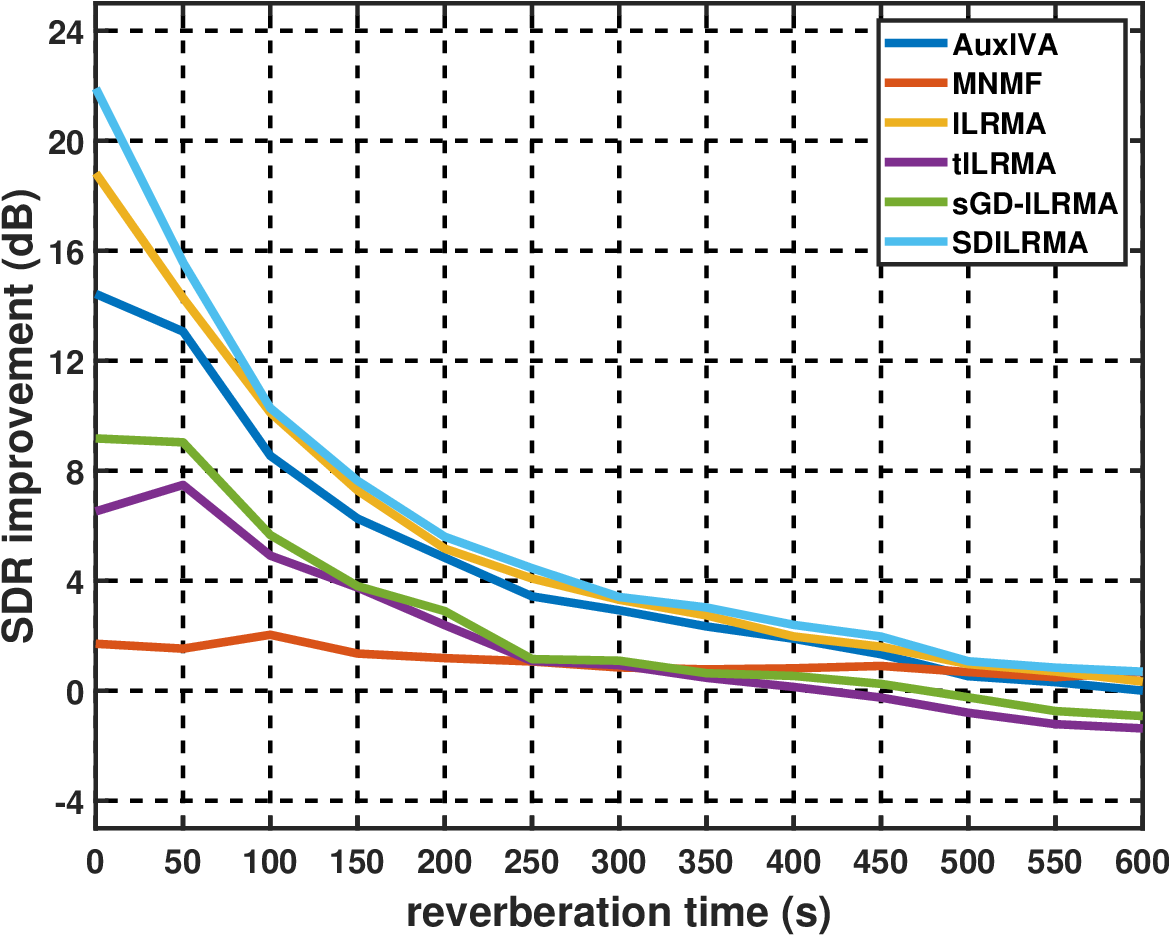}%
\label{fig_SISEC2011_5a}}
\hfil
\subfloat[female+male ]{\includegraphics[width=1.85in]{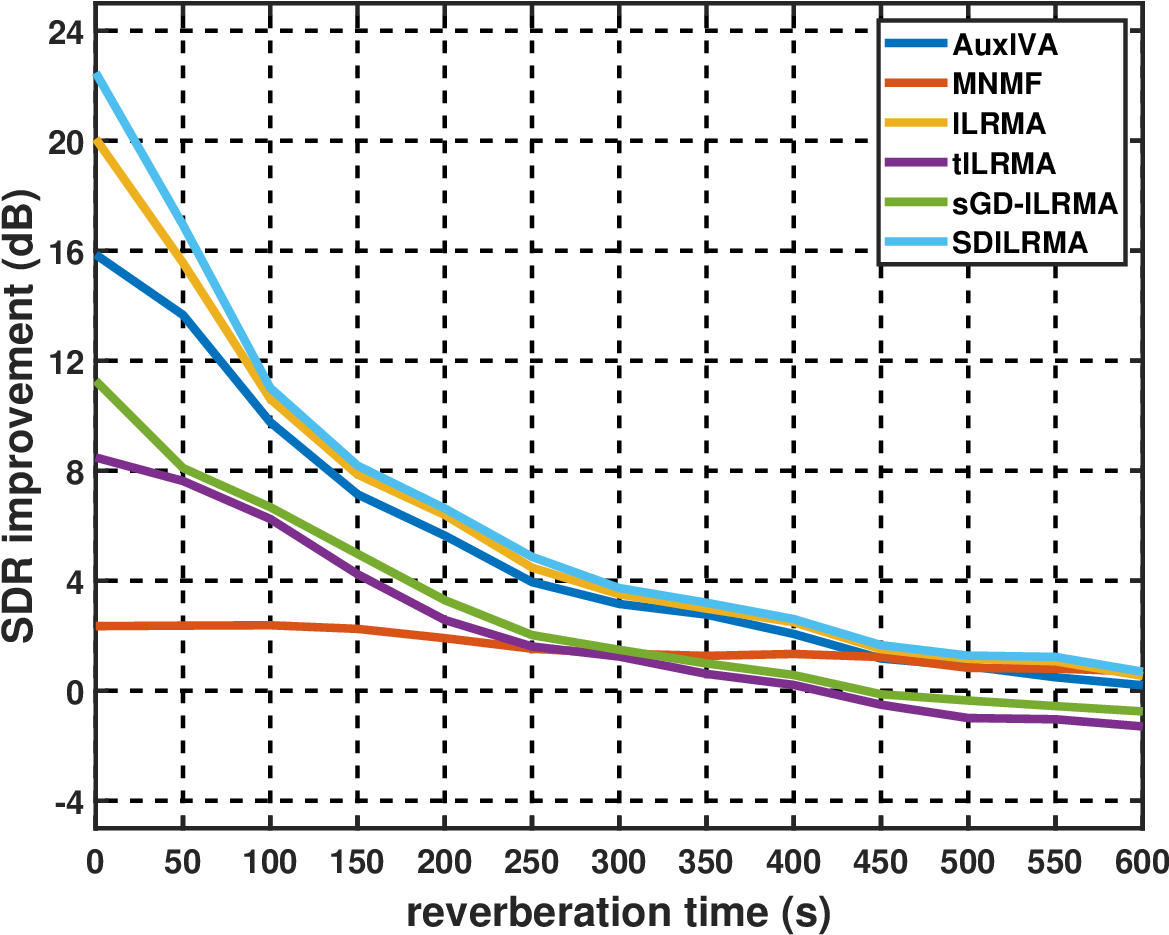}%
\label{fig_SISEC2011_5b}}
\hfil
\subfloat[male+male ]{\includegraphics[width=1.85in]{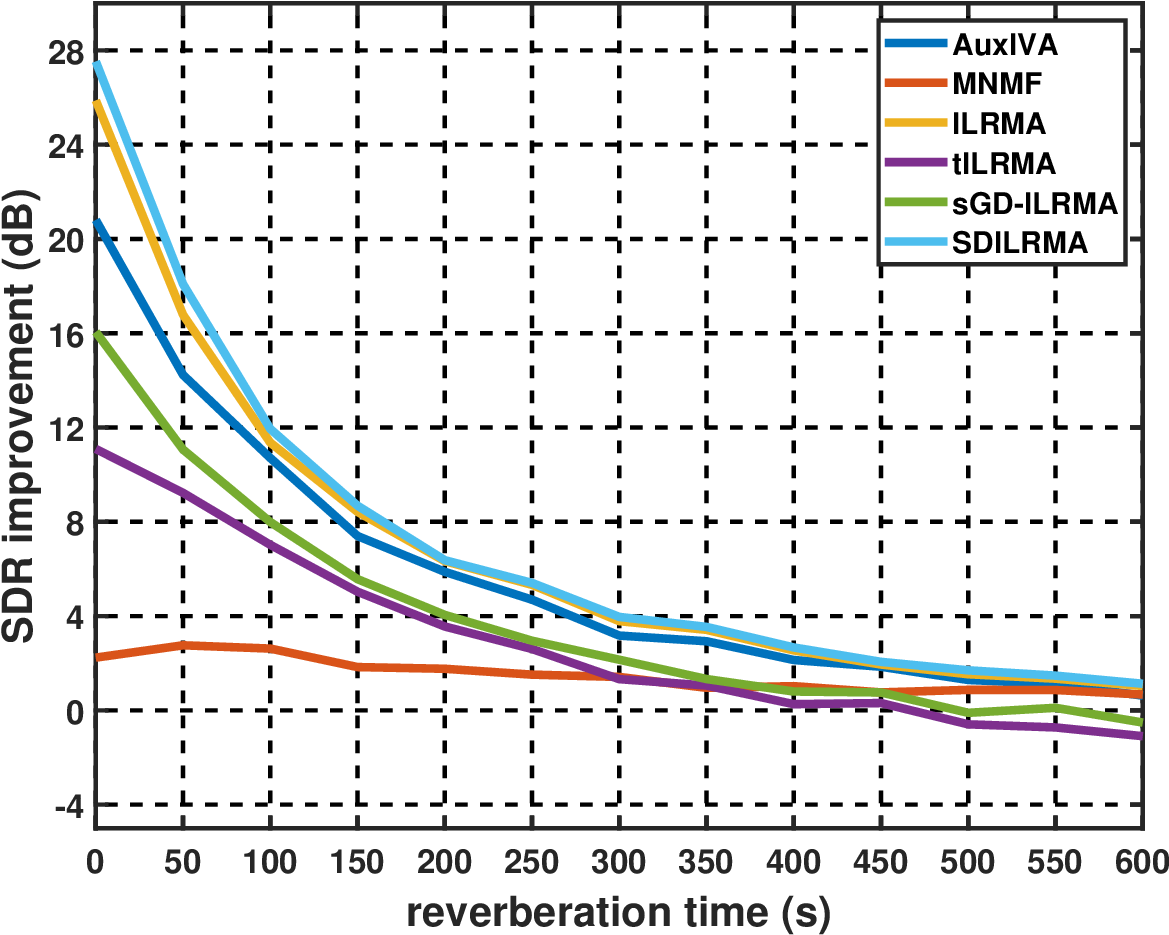}%
\label{fig_SISEC2011_5c}}
\hfil
\subfloat[female+female ]{\includegraphics[width=1.85in]{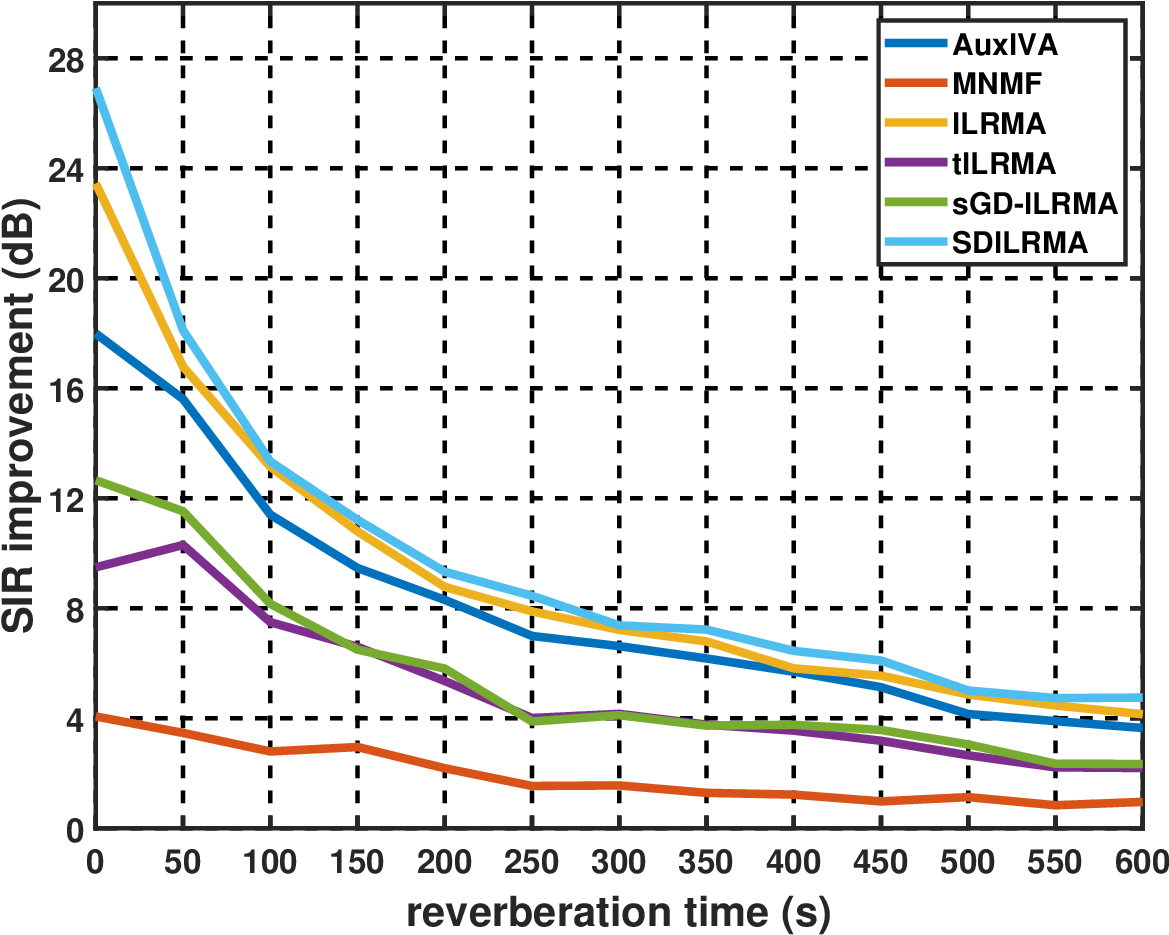}%
\label{fig_SISEC2011_5d}}
\hfil
\subfloat[female+male ]{\includegraphics[width=1.85in]{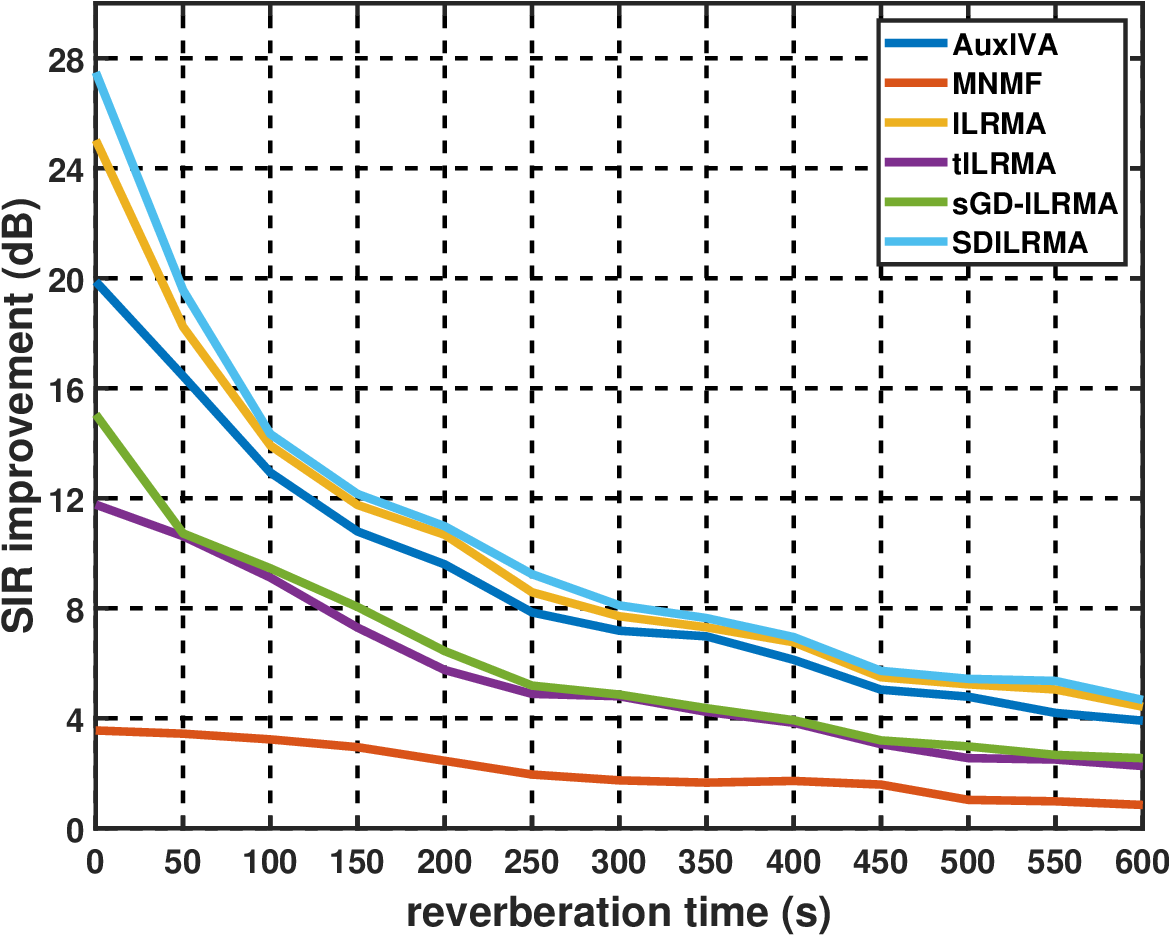}%
\label{fig_SISEC2011_5e}}
\hfil
\subfloat[male+male ]{\includegraphics[width=1.85in]{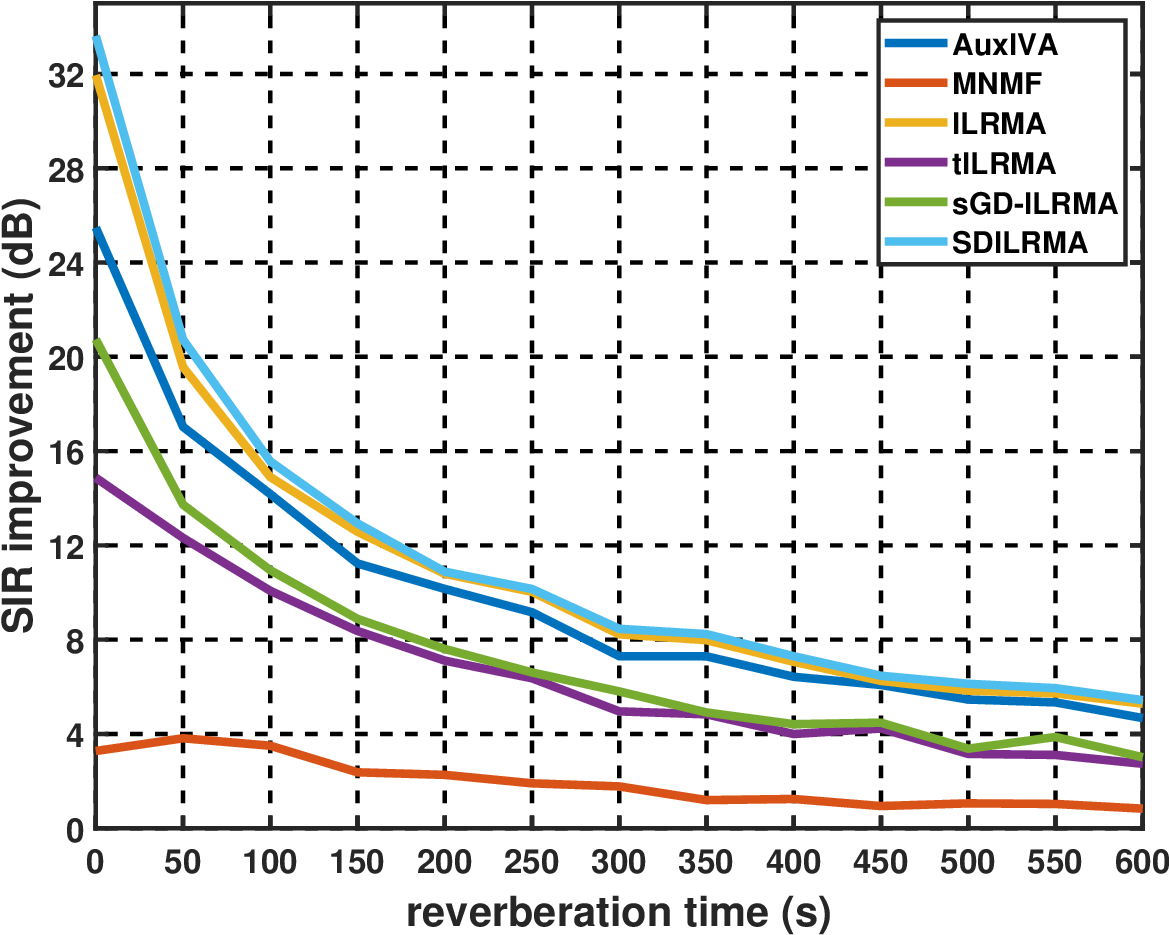}%
\label{fig_SISEC2011_5f}}
\caption{SDR and SIR improvement of the studied methods.}
\vspace{0.2cm}
\label{Fig2}
\end{figure*}

\section{SIMULATIONS} \vskip 6pt

We used some speech signals from the Wall Street Journal (WSJ0) corpus \cite{ref21} as the clean speech source signals and configured evaluation signals following the SISEC challenge \cite{ref23} with $M=N=2$, where the room size is $8 \times 8 \times 3$ m. The two sources are assumed to be $2$ m away from the center of the two microphones and the microphone spacing is $5.66$ cm. The incidence angles of the two sources are randomly selected from $[0^{\circ},90^{\circ}]$ and $[0^{\circ},-90^{\circ}]$ respectively per mixture, where the direction normal to the line connecting two microphones is $0^{\circ}$. The image source model \cite{ref25} is used to generate the room impulse responses, where the sound absorption coefficients are calculated by Sabine's Formula \cite{ref26} with the room aforementioned room size and reverberation time $T_{60}$ changing from $0$ to $600$ ms with an interval of $50$ ms. For each combination of sources (there are four combinations) and every value of $T_{60}$, $100$ mixtures are generated for evaluation. The sampling rate is $16$ kHz.

The parameters $\mu$ and $\gamma$ of SDILRMA were set to $100$, and $10$, respectively. We compared SDILRMA with AuxIVA \cite{ref8}, MNMF \cite{ref22}, ILRMA \cite{ref5}, $t$-ILRMA and sGD-ILRMA \cite{ref20}. The performance metrics used are the signal-to-distortion ration (SDR) and source-to-interferences ratio (SIR) \cite{ref24}.

Figure~\ref{Fig2} presents the results in terms of the average SDR and SIR improvements. It is seen that SDILRMA outperforms MNMF, ILRMA, $t$-ILRMA and sGD-ILRMA, which demonstrates the effectiveness of SDILRMA for source separation.
\begin{figure}[t]\label{Fig3}
\vspace{-0.5cm}
\begin{minipage}[b]{1\linewidth}
  \centering
  \centerline{\includegraphics[width=16cm]{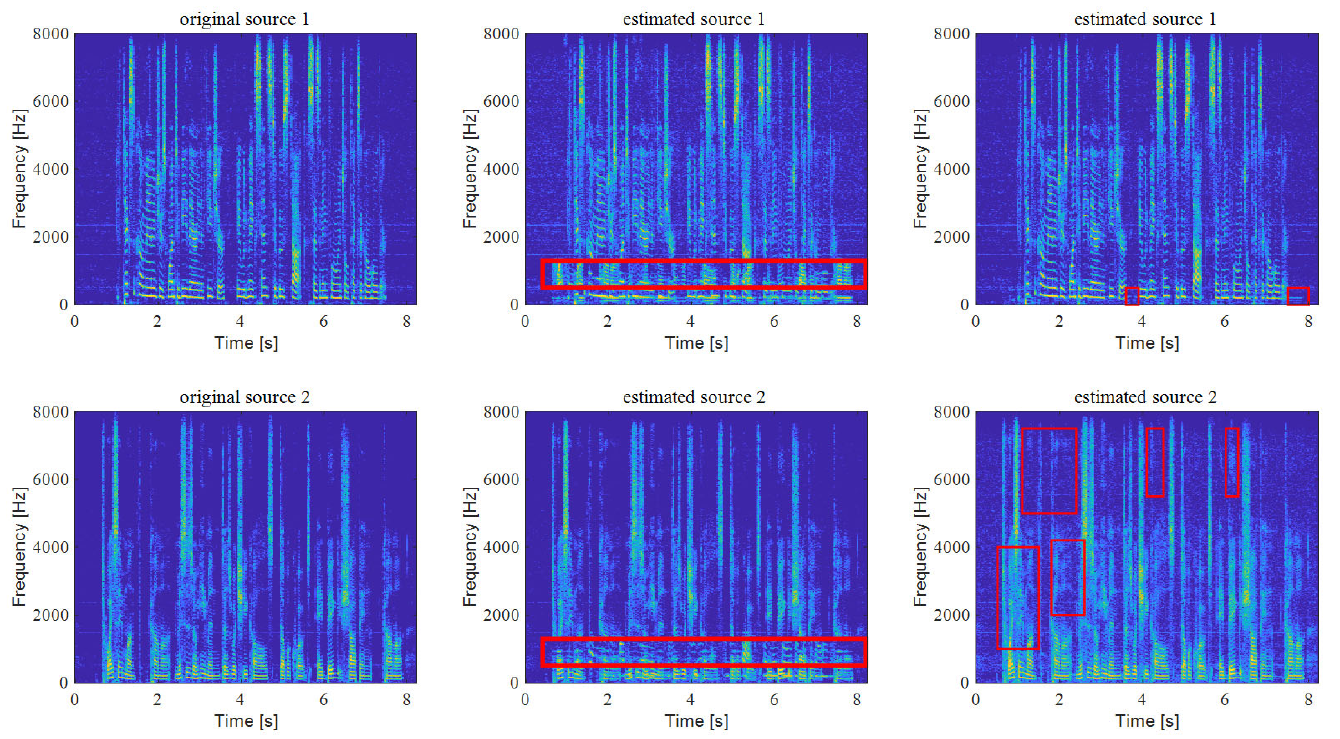}}
\end{minipage}
\caption{The spectrograms of  the source and separated signals. Left panels: the original source signals. Middle panels: the separated signals by ILRMA. Right panels: the separated signals by SDILRMA.}
\label{Fig3}
\end{figure}

Figure\ref{Fig3} plots the spectrograms of the source signals as well as the signals estimated by ILRMA and SDILRMA. It is seen that both ILRMA and SDILRMA are effective. ILRMA suffers from a small number of permutations, which are not seen in SDILRMA. This, again, demonstrates the superiority of 
SDILRMA.

\section{CONCLUSION} \vskip 6pt

This paper studied the determined BSS problem for audio and speech applications. We presented an improved version of ILRMA, which
applies NMF to decompose the time-varying source model and Sinkhorn divergence as the cost function to optimize the model parameters. To simplify the algorithm to reduce its computational complexity, the Kronecker product tool was used to decompose the modeling matrix into the product of a number of matrices of much smaller dimensionality, resulting in a simplified SDILRMA algorithm. Simulation results verified that the simplified SDILRMA is able to achieve better BSS performance than ILRMA and is also computationally more efficient than its counterpart without Kronecker product decomposition.

\bibliographystyle{abbrv}

\begin{thebibliography}{}

\end{thebibliography}


\noindent \section*{\uppercase{References}}
\vspace{-18pt}
\begin{thebibliography}{10}
\footnotesize
\vskip 6pt

\bibitem{ref1}
Benesty J, Makino S, Chen J. Speech enhancement. New York, NY, USA: Springer, 2005.

\bibitem{ref2}
Makino S, Lee TW, Sawada H. Blind Speech Separation.
New York, NY, USA: Springer, 2007.

\bibitem{ref3}
Li Y, Amari S, Cichocki A, Ho DWC, Xie S. Underdetermined blind source separation
based on sparse representation. IEEE Trans. Signal Process., vol. 54, no. 2, pp. 423--437, Feb. 2006.

\bibitem{ref4}
Bofill P, Zibulevsky M. Underdetermined blind source separation
using sparse representations. Signal Process. vol. 81, no. 11, pp. 2353--2362, Nov. 2001.

\bibitem{ref5}
Kitamura D, Ono N, Sawada H, Kameoka H, Saruwatari H. Determined blind source
separation unifying independent vector analysis and
nonnegative matrix factorization. IEEE/ACM Trans. Audio, Speech, Lang.
Process., vol. 14, no. 9, pp. 1626--1641, Sep. 2016.


\bibitem{benesty2011}
Benesty J, Chen J, Habets E. Speech Enhancement in the STFT Domain. Berlin: Springer-Verlag, 2011.

\bibitem{huang2014}
 Huang H, Zhao L, Benesty J, Chen J. A minimum-variance-distortionless-response
 filter based on the bifrequency spectrum for single-channel noise
 reduction. Digital Sig. Process.,  vol. 33, pp. 169--179, Oct. 2014.

\bibitem{ref6}
Comon P. Independent component analysis, a new concept. Signal Process., vol. 36, no. 3, 
pp. 287--314, Apri. 1994.

\bibitem{ref7}
Kim T, Eltoft T, Lee TW. Independent vector analysis: An extension of ICA to multivariate components.
in Proc. Int. Conf. Independent Compon. Anal. Blind Source Separation, Oct. 2006, pp. 165--172.


\bibitem{ref8}
Ono N. Stable and fast update rules for independent vector analysis
based on auxiliary function technique. in Proc. IEEE Workshop Appl.
Signal Process. Audio Acoust., 2011, pp. 189--192.

\bibitem{ref9}
Vallender S. Calculation of the wasserstein distance between probability distributions on the line.
Theory of Probability\& Its Applications, vol. 18, no. 4, pp. 784--786, 1974.


\bibitem{ref10}
Kantorovic LV, Rubinstein GS. On a functional space and certain extremum problems.
in \emph{Doklady Akademii Nauk},  Russian Academy of Sciences, vol. 115, pp.
  1058--1061, 1957.


\bibitem{ref11}
Cuturi M. Sinkhorn distances: Light-speed computation of optimal transport. in Advances in Neural Information
Processing Systems (NIPS). Red Hook, NY, USA: Curran, 2013, pp. 2292–-2300.


\bibitem{ref12}
Rolet A, Cuturi M, Peyré G. Fast dictionary learning with a smoothed wasserstein loss. in Proc. Int. Conf.
Artif. Intell. Stat., Cadiz, Spain, May 2016, pp. 630–-638.


\bibitem{ref13}
Benzi M, Simoncini V. Approximation of functions of large
matrices with Kronecker structure. Numerische Mathematik, vol. 135 no. 1,  pp. 1–-26, Jan. 2017.

\bibitem{ref14}
Motamed M. Hierarchical Low-Rank Approximation of Regularized Wasserstein Distance. arXiv preprint arXiv:2004.12511, 2020.


\bibitem{ref15}
Févotte C, Cardoso JF. Maximum likelihood approach for blind audio source separation using time-frequency Gaussian source models. in Proc. IEEE Workshop Appl. Signal Process. Audio Acoust., Oct. 2005, pp. 78--81.

\bibitem{ref16}
Vincent E, Jafari MG, Abdallah SA, Plumbley MD, Davies ME. Probabilistic modeling paradigms for audio source separation. In Machine Audition: Principles, Algorithms and Systems. IGI global, pp. 162--185, 2011.


\bibitem{ref17}
Brehm H, Stammler W. Description and generation of spherically invariant speech-model signals. Signal Process. 
vol. 12, no. 2, pp. 119--141, Mar. 1987.


\bibitem{ref18}
Buchner H, Aichner R, Kellermann W. Blind source separation for convolutive mixtures: A unified treatment.
In Audio signal processing for next-generation multimedia communication systems (pp. 255--293).
Springer, Boston, MA, 2004.

\bibitem{ref19}
Wang J, Guan S, Liu S, Zhang XL. Minimum-Volume Multichannel Nonnegative Matrix Factorization for Blind Audio Source Separation. {IEEE/ACM Trans. Audio, Speech, Lang. Process.}, vol.~23, no.
  3089--3103, Aug. 2021.


\bibitem{ref20}
Mogami S, Takamune N, Kitamura D, Saruwatari H, Takahashi Y, Kondo K, Ono N. Independent low-rank matrix analysis based on time-variant sub-Gaussian source model for determined blind source separation. {IEEE/ACM Trans. Audio, Speech, Lang. Process.}, vol.~28, pp. 503--518,
 Dec. 2019.

\bibitem{ref21}
Garofolo J, Graff D, Paul D, Pallett D. Csr-i (wsj0) complete ldc93s6a. Web Download. Philadelphia:
Linguistic Data Consortium, 83, 1993.


\bibitem{ref22}
Sawada H, Kameoka H, Araki S, Ueda N. Multichannel extensions of non-negative matrix factorization
with complex-valued data. IEEE Trans. Audio Speech Lang. Process, vol. 21, no. 5, pp. 971--982, Jan. 2013.


\bibitem{ref23}
Araki S, Nesta F, Vincent E, Koldovský Z, Nolte G, Ziehe A, Benichoux A. The 2011 signal separation
evaluation campaign (SiSEC2011):-audio source separation. LVA/ICA (pp. 414--422). Springer, Berlin,
Heidelberg 2012.

\bibitem{ref24}
Vincent E, Gribonval R, Févotte C. Performance measurement in blind audio source separation. {IEEE Trans. Audio, Speech, Lang.
  Process.}, vol.~14, no.~4, pp. 1462--1469, June 2006.

\bibitem{ref25}
Allen JB, Berkley DA. Image method for efficiently simulating small-room acoustics. J. Acoust. Soc. Amer.,
vol. 65, no. 4, pp. 943--950, Apr. 1979.

\bibitem{ref26}
Young RW. Sabine reverberation equation and sound power calculations. The Journal of the Acoustical
Society of America, vol. 31, no. 7, pp. 912–-921, July 1959.

\end{thebibliography}
\renewcommand{\refname}{\normalfont\selectfont\normalsize}
 
\end{document}